# Ab-initio calculations of the Optical band-gap of $TiO_2$ thin films

Wilfried Wunderlich[1*], Lei Miao[1], Masaki Tanemura[1], Sakae Tanemura[1], Ping Jin[2],
Kenji Kaneko[3], Asuka Terai[4], Nataliya Nabatova-Gabin[4] and Rachid Belkada[5]

[1]Nagoya Institute of Technology, Dept. of Environmental Engineering, Showaku, 466-8555 Nagoya
[2]Chubu Centre, Dept. Metal & Inorganic Material, National Institute of Advanced Industrial Science & Technology, Moriyama-ku, Nagoya 463-8560, Japan
[3]Dept. Material Science and Engineering & HVEM Lab., Higashi-ku, Fukuoka 812-8581, Japan
[4]Horiba Jobin Yvon Co. Ltd., Chiyoda-ku, Tokyo 101-0031, Japan
[5]Japan Science and Technology Corporation, Kawaguchi, 332-0012, Japan
* corresponding author, email: wunder@system.nitech.ac.jp

**Abstract**
Titanium dioxide has been extensively studied in recent decades for its important photocatalytic application in environmental purification. The search for a method to narrow the optical band-gap of $TiO_2$ plays a key role for enhancing its photocatalytic application. The optical band gap of epitaxial rutile and anatase $TiO_2$ thin films deposited by helicon magnetron sputtering on sapphire and on $SrTiO_3$ substrates was correlated to the lattice constants estimated from HRTEM images and SAED. The optical band-gap of 3.03 eV for bulk-rutile increased for the thin films to 3.37 on sapphire. The band gap of 3.20 eV for bulk-anatase increases to 3.51 on $SrTiO_3$. In order to interpret the optical band gap expansion for both phases, ab-initio calculations were performed using the Vienna ab-initio software. The calculations for rutile as well anatase show an almost linear increase of the band gap width with decreasing volume or increasing lattice constant *a*. The calculated band gap fits well with the experimental values. The conclusion from these calculations is, in order to achieve a smaller band-gap for both, rutile or anatase, the lattice constants *c* has to be compressed, and *a* has to be expanded.

## 1. Introduction

Titania in both modifications, anatase or rutile phase, is one of the most important material for applications based on photon excitation. The efficiency of heterogeneous photocatalytic devices [1] or photovoltaic solar cells [2] can be improved, when the band gap of the light sensitive titania layer can be narrowed. The optical band gap of such thin films was estimated [3-10] by spectroscopic ellipsometry (SE), where the optical properties like the extinction coefficients *k* of the thin film are measured as a function of the photon energy followed by an appropriate standard analyzing method according to Tauc [11]. The challenge for material scientist is to find a processing method in order to realize the smaller band gap, since it strongly depends on the microstructure obtained from the different processing conditions as described in the following. Epitaxial thin film growth is known as a suitable method to change the lattice constants of thin films by matching to those of a suitable substrate. When the thickness during growth reaches a critical thickness, misfit dislocations can be formed [12,13]. Epitaxial titania thin films without misfit dislocations have been produced by different sputtering methods [3-8], while thin films prepared by sol-gel method [9-11] have a less dense structure. In all of these investigations, the band gap value of the thin films is usually larger than the bulk values [14], 3.03eV for rutile and 3.20eV for anatase, as shown in table 1 and can be correlated to the lattice constants, measured by X-ray diffraction, electron diffraction or the lattice fringe spacing in HRTEM micrographs. the results from the literature. Furthermore, nano-crystalline Titania shows a wider band gap [15], while nano-porous Titania in both phases has a smaller band gap, 3.15 or 2.55eV [16], repetively.

In order to understand the change of the band gap value, the density of states and the energy of the electronic orbitals need to be known. Ab-initio simulations based on density functional theory (DFT) for calculating the electronic wave functions and their periodicity in the crystal are required. Such a software with high reliability and reputation is the Vienna Simulation Program (Vasp) [17], which has been successfully applied for anatase and rutile [18-19] for calculating the surface energy and adsorption. The electronic band structure of rutile and anatase has been calculated by similar ab-initio methods [20-21] and the benefit of such simulations for material processing will become even more important in future [22]. This paper reports about ab-initio calcaultaions closely related to the above mentioned experimental results, about band gap values related to the lattice constants and their interpretation by ab-initio calculations.

## 2. Calculation method

According to the experimentally obtained data the unit cells of rutile or anatase thin films on substrates are distorted. For modeling this behavior, the supercells for simulation were constructed by assuming the same atomic position as in the single crystal, but expansion or shrinkage of the unit cells. The Vienna ab-initio simulation package (VASP) [16-18] was used for calculating the density of states (DOS) of anatase and rutile. This software is based on the density functional theory (DFT) by solving the Kohn-Sham modified Schroedinger equation with the main assumption, that the wave function of a single electron can be treated independent from other electrons. This effective calculation method treats the contribution from core electrons by ultra-soft pseudo potentials, which are available for most elements, including titanium and oxygen atoms, from the Vasp library. Vasp allows calculations of density of states, and the orbital band structure with high precision. Necessary checks are the sufficient number of k-points, and the size of the cut-off energy. Finally it was proofed, that the total energy versus volume E(V)- shows almost the same equilibrium lattice constants as the literature data for both phases anatase and rutile.

The density of states (DOS) was plotted and the gap between the oxygen 2p-states and the titania 3d-states was measured, by assuming a threshold value of 0.3 of the maximum of the DOS. This value was obtained from calibrating the band gap to the experimental value of the band gap for bulk-rutile of 3.03eV. Using this method, the band-gap of bulk-anatase was obtained as 3.20eV, which is in excellent agreement to the experiments. This calibration method uses the *ad-hoc* assumption, that a certain density of electrons is required to show an effect in the experiment, and is justified by its success, because the complete understanding of the complicated excitation and recombination process of electrons in certain orbitals by photons with certain energy is still a challenging field of solid-state physics. The stress-condition of the thin film on the substrate was modeled using supercells with different lattice constants $a$, $c$ and $a/c$ ratios by varying $a$ and $c$ in the range from 0.92 to 1.02 of the bulk values for both, anatase and rutile. All calculations were performed under the same conditions, allowing the comparison between them.

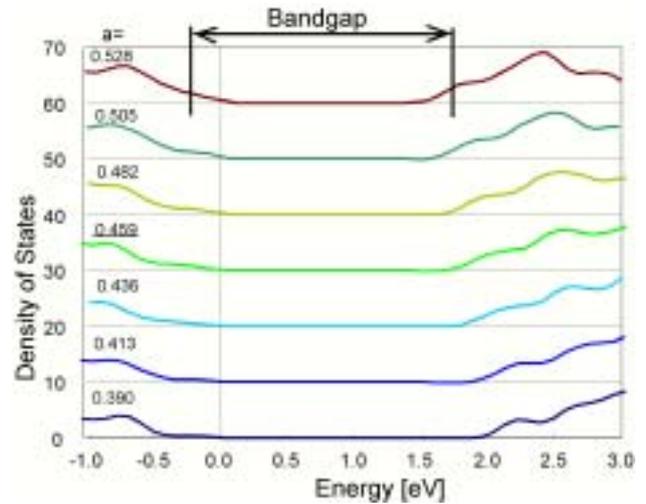

Fig. 1 Estimation of the band gap for rutile from the plot of the Density of states versus the Energy for different values of the lattice parameter a in [nm]

## 3. Results and Discussion

The experimental data showed a correlation between the lattice distortion and the band gap width, and the goal of the ab-initio simulations is to understand this behavior. The strain or compression of the *a*- or *c*-lattice constant is used in the simulation to model the distortion of the crystal lattice of the thin film on the substrate. The density of states as a function of the energy is shown in Fig. 1, where the lattice constant *a* for rutile is varied compared to the bulk value of *a*=0.459nm as marked on the left side. The band gap was estimated as marked after proper calibration as described above. The band gap increases when the lattice constant *a* is decreased.

Table 1) Experimental results of the correlation between lattice constants and optical band gap of anatase and rutile thin films

| Substrate | Orientation Relationship S=Substrate, An=Anatase, Ru=Rutile | Lattice constant a [0.1nm] | c [0.1nm] | Band gap [eV] | Literature reference |
|---|---|---|---|---|---|
| Anatase -bulk | | 3.785 | 9.514 | 3.20 | [14] |
| Rutile -bulk | | 4.593 | 2.959 | 3.03 | [14] |
| SrTiO$_3$ | (100)_S//(100)_An, [001]_S // [001]_An | 3.660 | 9.760 | 3.51 | [3] |
| α-Al$_2$O$_3$ | (11.0)_S//(100)_Ru, [00.1]_S//[010]_Ru | 4.460 | 3.120 | 3.37 | [3] |

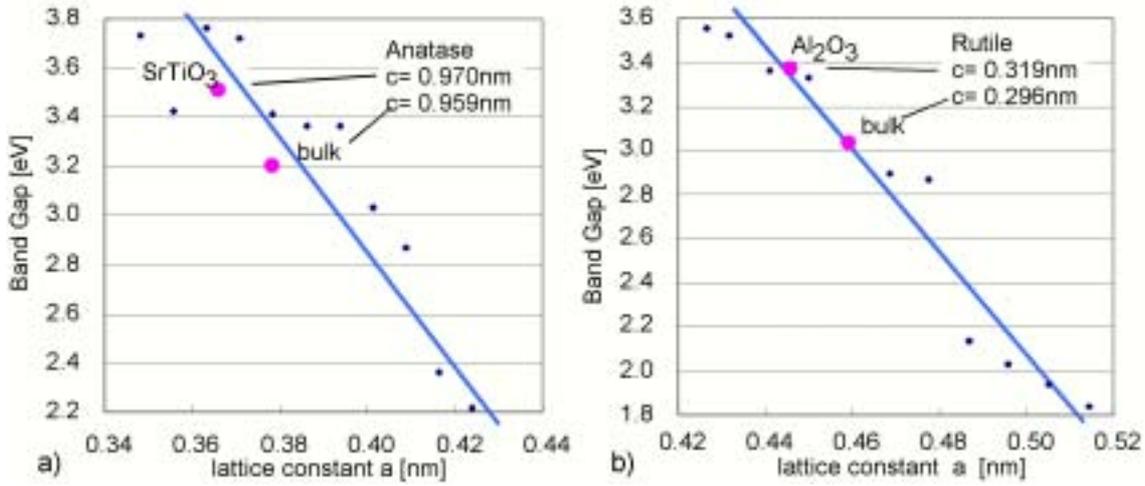

Fig. 2 Comparison between experimental data (large circles) and calculation data of the band-gap as a function of the lattice constant a in [nm] for a) Anatase b) Rutile

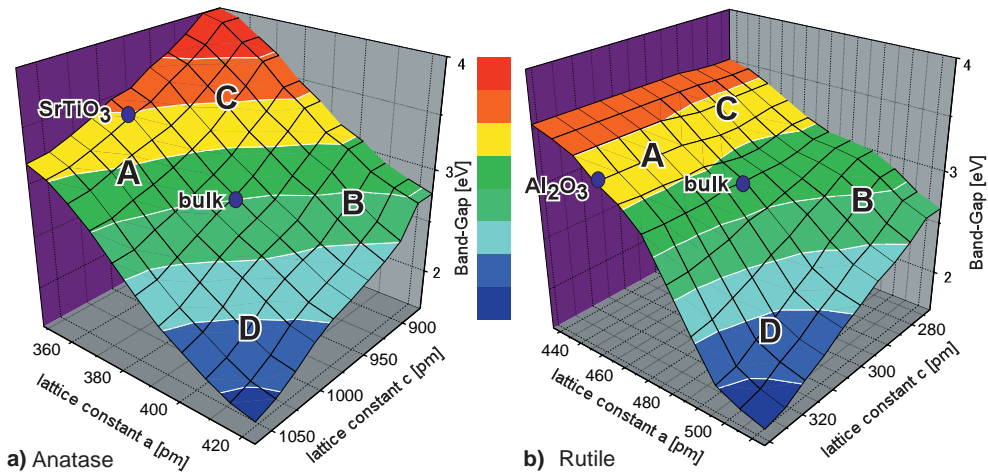

Fig. 3 Calculated Band gap width in [eV] as a function of the lattice constants *a* and *c* in [pm=0.001nm] for a) Anatase and b) Rutile. The dots mark the experimental data points for bulk and thin films on $Al_2O_3$ or $SrTiO_3$. Refer the text for the regions marked with letters A-D.

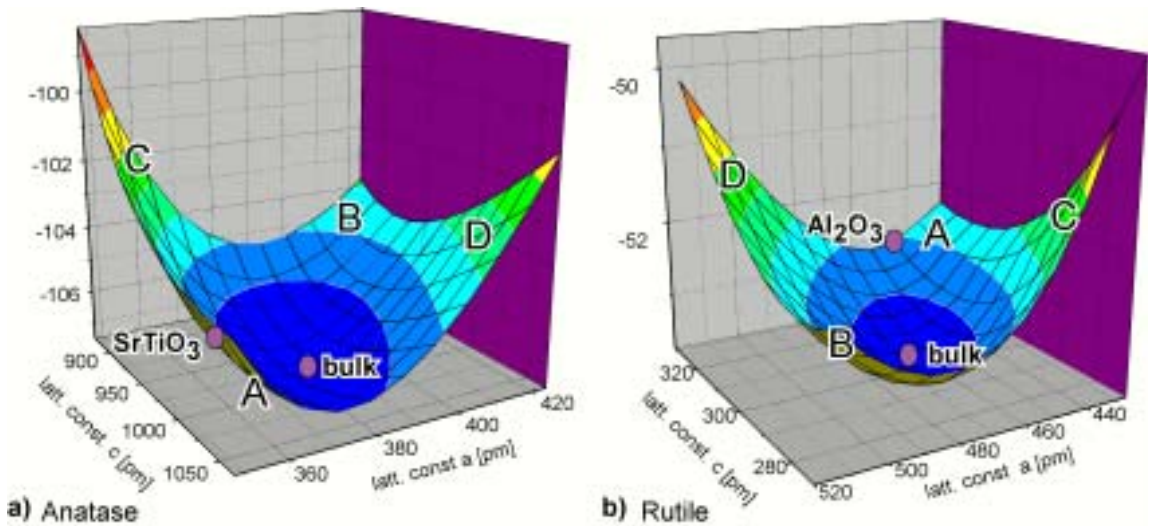

Fig. 4 Total energy of the simulations as a function of the lattice constants for a) Anatase and b) Rutile. The dots mark the experimental data and the regions A-D correspond to those in fig. 6.

The calculations were performed for a wide range of variations of *a* and *c* for both phases, anatase and rutile. Fig. 2. shows that the comparison between experimental data shown as large dots and calculated data is excellent dependence for both, the thin film as well as the bulk values, for both phases a) anatase and b) rutile. The data were plotted in the three-dimensional graph in Fig. 3, in order to show the dependence of the band-gap (z-axis) on both lattice constants a and c (x- and y-axis). In both cases, a) anatase and b) rutile, the band gap decreases with increasing a and c, but the dependence on c is less for the rutile phase. The values for the thin films lie in both cases in region A, which means a slight increase in the band gap. The desired decrease in the band gap to values of less than 2eV would occur in region B.

The band gap estimations from DOS calculations show very good agreement to the experimental values as can be seen in Fig. 2. The bulk values for anatase and rutile and those for the epitaxial thin films on two different two substrates agree very well with the average line of the calculated values. The discrete nature of the simulations for lattice constant, k-point spacing and electron waves can explain the scattering in the calculation data. The band gap increases due to expansion of the lattice constant *c*, while the lattice constant *a* has a smaller value than the bulk, in agreement with the experimental results and literature values [3-10]. The calculation results show, that the desired decrease of the band gap requires the opposite, expansion in *a*-direction and shrinkage in *c*-direction. Whether this can be achieved in the experiment is doubtful, as shown in the following Fig. 4, which shows the lattice energy of a) the anatase, b) rutile phase as a function of the lattice parameters *a* and *c*. For clarification the axis of this 3-d plot are rotated compared to fig. 3, and the corresponding regions A-D are marked. It is obvious, that the lattices of both phases with the bulk lattice constants (dot in the middle) have the lowest energies, showing the excellent performance and reliability of the ab-initio software Vasp. Both thin film lattices have energies only slightly larger than the bulk lattice, located in region A. The desired low-band gap region D has a much higher energy unlikely to be reached in stable films, while region B with a moderate decrease in the band gap shows about the same increase in energy as the present thin films. The results can be explained by the volume conservation law, which states that an increase in one of the lattice constants should be compensated by a decrease in the other one, and vice-versa, as it is the case in the low-energy regions A and B. Hence, it seems to be possible, that this decrease in the band gap can be achieved by further search for a suitable substrate, on which the optimal deformation state with elongated *a*- and compressed *c*-axis for thin films growth is realized. These calculation results strongly recommend to consider the application of additional stress on the substrate during or after processing, which has been successful in optimizing many other interfaces [12]. Further studies are in progress, in order to clarify the influence of the size-effect of crystals in nano-meter dimensions or nano-porous materials.

**Summary**

This study concerned the structural and optical properties of $TiO_2$ thin films prepared by rf magnetron sputtering under a precise control of the growth parameters. The correlation between lattice constants of the thin films and the optical band gap showed an increase compared to the bulk values for both, anatase and rutile. This behavior, 3.37eV on a sapphire (11.0) substrate compared to 3.03 eV for bulk-rutile, and 3.51eV on SrTO3 compared to 3.20 eV for bulk-anatase, can be verified by the ab-initio simulations, which are in excellent agreement. These calculations allow the predictions, that under certain stress conditions the desired narrowing of the anatase or rutile band-gap can be achieved.